\documentclass[%
 aip,
 jap,%
 amsmath,amssymb,
 reprint,%
]{revtex4-1}

\usepackage{graphicx}% Include figure files
\usepackage{dcolumn}% Align table columns on decimal point
\usepackage{bm}% bold math
\begin{document}

\preprint{}

\title[Design of a 4-electrode optical device for application of vector electric fields to self-assembled quantum dot complexes]{Design of 4-electrode optical device for application of vector electric fields to self-assembled quantum dot complexes}% Force line breaks with \\

\author{Xinran Zhou}
\author{Matthew Doty}%
 \email{doty@udel.edu.}
\affiliation{
University of Delaware, Newark, DE 19716, USA%\\This line break forced with \textbackslash\textbackslash
}%

\date{\today}% It is always \today, today,

\begin{abstract}
Self-assembled InAs quantum dots (QDs) are of great interest as components of optoelectronic devices that can operate at the quantum limit. The charge configuration, interdot coupling, and symmetry of complexes containing multiple QDs can all be tuned with applied electric fields, but the magnitude and angle of the electric field required to control each of these parameters depends on the orientation of the QD complex. We present a 4-electrode device compatible with optical excitation and emission that allows application of electric fields with arbitrary magnitudes and angles relative to isolated QD complexes. We demonstrate the electric field tunability of this device with numerical simulations.
\end{abstract}

%Use showkeys class option if keyword
\maketitle

Self-assembled quantum dots (QDs) are of great interest for next-generation optoelectronic devices such as entangled photon emitters/ detectors and quantum information processors.\cite{Suwit2009, Muller2014} Such devices are anticipated to have new functionality arising from the unique quantum mechanical properties of the QDs, which include discrete charge states, well-defined spin projections and outstanding optical efficiencies. Although the size and composition of InAs QDs can be used to tailor the wavelength of QD optical transitions during growth, the self-assembly process that leads to QD formation also leads to an unavoidable inhomogeneous distribution of the transition wavelengths. Methods to tune the wavelength of QD optical transitions \emph{in-situ} are thus of great importance. One way to tune the energies of single QDs is to utilize the Stark shift caused by local electric fields.\cite{Kaniber2011} Another method to tune the emission wavelength of discrete states is to make use of the indirect transitions in complexes of QDs, also known as quantum dot molecules (QDMs).\cite{Wang2009} Indirect transitions in QDMs can have wavelengths that tune strongly with applied electric field because they involve electrons and hole predominantly located in separate QDs.

The nature and strength of optical transitions, Coulomb interactions and exchange interactions in QDMs depend on the orientation of the QDs relative to one another and the growth axis of the heterostructure.\cite{Stinaff2006, Doty2009, Liu2013, Zhou2013} The inter-dot coupling of two QDs aligned along the growth (vertical) axis is typically relatively strong because truncated QDs have a pancake shape and the center-to-center separation along the growth direction can be quite small. Side-by-side (lateral) QDMs, on the other hand, typically have a weaker tunnel coupling because the center-to-center distance is larger.\cite{Lee2010, Zhou2013} Both types of QDMs can be controllably charged by application of electric fields along the growth direction, which tune the confined energy levels of the QDs relative to the Fermi level set by a doped substrate. In the case of vertical QDMs, such a vertical electric field controls \emph{both} the total charge occupancy and the relative energy levels of the two QDs, which controls the tunnel coupling and the formation of molecular states with unique properties. Application of electric fields with lateral components to vertical QDMs could be used to break the molecular symmetry, which is known to enable new spin mixing phenomena with important technological applications.\cite{Doty2010a, Economou2012} In the case of lateral QDMs (LQDMs), the charge occupancy and tunnel coupling could be controlled \emph{independently} by electric fields along the growth (vertical) and molecular (QD-to-QD) axes, respectively.\cite{Zhou2011} In the case of single QDs, tunable vector electric fields would enable analysis and exploitation of exciton fine-structure splitting by manipulating the wavefunction overlap of electrons and holes in both the vertical and lateral dimensions.\cite{Mar2010, bennett2010nat}
\begin{table*}[htbp]
\centering
\caption{Design criteria for vector electric field device incorporating LQDMs}
\begin{tabular}{ccc}\hline
Property & Target Value & Motivation\\\hline
Aperture size & $\sim$1 $\mu$m & Isolate single QDMs\cite{Zhou2011} at densities of $\sim30/\mu m^2$ \\ & & Allow optical measurements with lasers of 800-1000 nm wavelength\\ & & Allow fabrication by optical lithography\\\hline
Distance from QDs to doped layer & $\leqslant$120 nm & Adequate charge tunneling rate\cite{Baier2001, Liu2013}\\\hline
Minimum lateral field uniformity  & $\geqslant$ 50 nm & Uniform lateral field over a LQDM\cite{Lee2010} \\\hline
Max lateral field & $\geqslant$0.5 MV/m & Tune the relative energy difference\\ & & between LQDMs by at least 10 meV\cite{Zhou2013}\\\hline
Max leakage current & $\leqslant$ 100 mA & Prevent excessive heat generation and sample damage\\\hline
\end{tabular}{}
\end{table*}

\begin{figure}[htbp]
\centerline{\includegraphics[width=6.5cm]{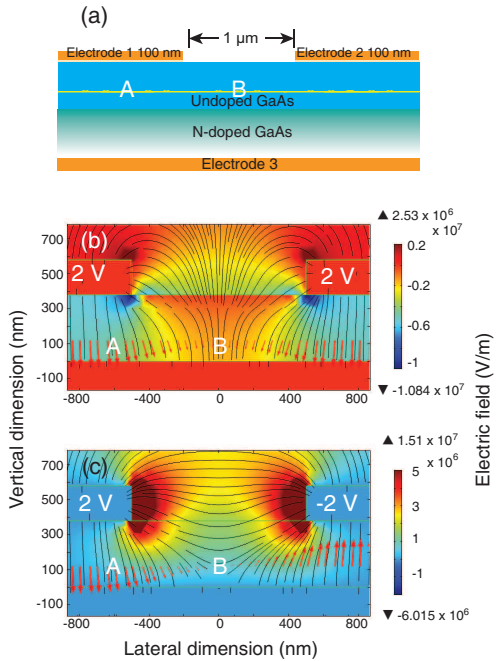}}
\caption{(Color Online) (a) Schematic diagram of 3-terminal device. (b,c) Electric field distributions in the 3-terminal device when applying (b) pure vertical or (c) pure lateral electric field as measured at the center of the aperture. The color shows the intensity of electric fields along (b) growth direction or (c) molecular axis; the electric field lines are shown in black; the arrows show the direction and magnitude of the electric field along the location of the LQDMs layer.\label{3ele}}
\end{figure}
Current QD device architectures apply electric fields in only one direction, along \cite{Liu2013, Zhou2011, Zhou2013, Liu2011, Mar2011} or perpendicular to \cite{Munoz2011, Hermannstadter2010, Kaniber2011, Vogel2007} the growth direction. Future optoelectronic devices would gain substantial functionality from the application of ``vector" electric fields that control both the magnitude and angle of the electric field applied to individual QDs or QDMs. Because controlled nucleation of InAs QDs at pre-determined locations remains challenging, one of the significant challenges for a ``vector" electric field device is to simultaneously control lateral and vertical electric fields for QDs or QDMs at arbitrary locations between the electrodes. Another challenge is to design electrodes that apply relatively uniform lateral electric fields to LQDMs that have a large lateral separation between the QDs. Ideally, the electrodes would also serve as an aperture that isolates single QDs or QDMs for quantum device applications.

This Letter introduces a device designed to apply ``vector" electric fields to single QDs or QDMs at arbitrary locations. We use LQDMs to illustrate the device geometry because application of a purely lateral electric field along the inter-QD axis of a LQDM at an arbitrary location within a mesa provides the most challenging device design conditions. We consider a LQDM with a QD center-to-center distance of 50 nm, which is typical of real LQDMs.\cite{Zhou2013} Table 1 summarizes the design criteria for the device.

The simplest way to fabricate a device that might apply vector electric fields to QDs or QDMs is to create a single back contact and two split electrodes on the top surface. Fig.~\ref{3ele}(a) shows the design for such a 3-electrode device. The advantage of the 3-electrode design is that it could be fabricated with only one photolithography and lift off step. However, as we will show, the performance of this device is severely limited. The top electrodes, separated by a 1 $\mu$m gap, serve as both lateral gates and an aperture for isolating single QDs. This device can apply a vertical electric field to the center of the aperture by applying voltages with the same magnitude and same sign (relative to the back contact) to the two electrodes on the top surface. This device can similarly apply a lateral electric field to the center of the aperture by applying voltages with the same magnitude and opposite sign to the two top electrodes. To illustrate the limitations of this design, we present device simulations generated with COMSOL Multiphysics of the uniformity and tunability of the electric fields that can be applied.

Fig.~\ref{3ele} (b,c) shows that the three-terminal device could apply a lateral and vertical electric field to a QD complex in the center of the aperture (point B). However, the fields that can be applied are highly nonuniform across the aperture, as shown by the differences between points A and B in Fig.~\ref{3ele}(b,c). The non-uniformity of the vertical electric field arises because of the relatively large lateral and vertical separation between the top electrodes and the center of the aperture. The vertical electric field applied at the center of the aperture (Point B of Fig.~\ref{3ele} (b)) is four times smaller than the electric field under the electrodes (Point A). As a result, most of the device, which is covered by the metal electrodes, will reach flat-band voltage and exceed the allowed leakage current before the electric field at point B is large enough to charge the QDs. Similarly, the non-uniform lateral field arises because of the back contact that cannot be at the same potential relative to both lateral contacts. Moreover, it is difficult to apply a uniform lateral field to QDs that are not located at the center of the aperture, a condition that is inevitable with present QD growth technologies. Resolving these challenges with only two top electrodes is prohibitive. A modified 3-electrode design (not shown) includes a pair of electrodes surrounding the QDs and on the same horizontal plane as the QD layer. Such a design can apply uniform lateral electric fields, but cannot apply vertical fields because the QDs are not covered by electrodes.
\begin{figure}[htbp]
\centerline{\includegraphics[width=8.0cm]{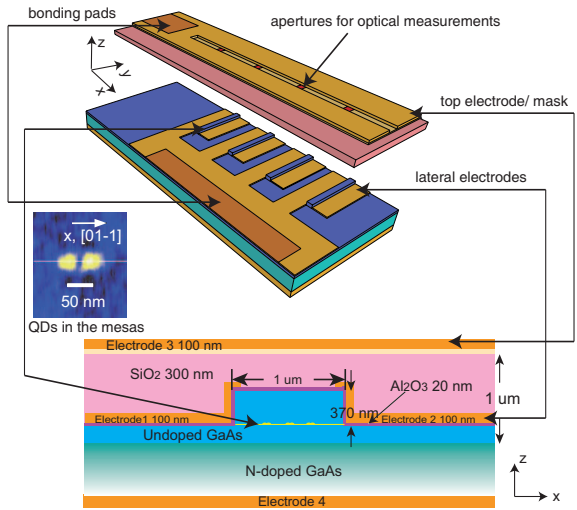}}
\caption{(Color Online) Schematic diagram of 4-electrode device design in perspective view (top) and cross-sectional view (bottom). An atomic force microscopy image of a single lateral QDM inside the mesa is shown in the inset.\label{4ele}}
\end{figure}

\begin{figure*}[htbp]
\centerline{\includegraphics[width=14.0cm]{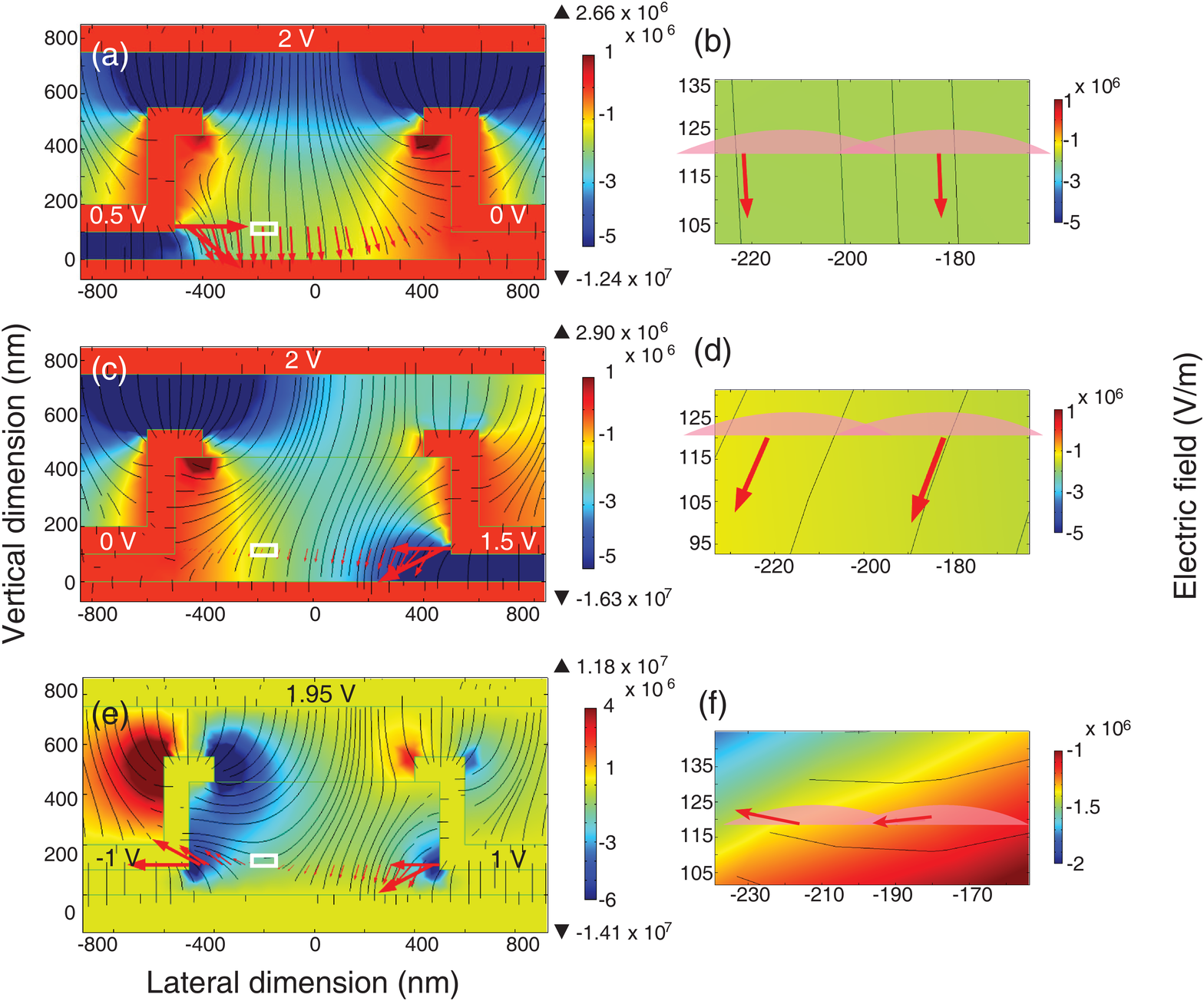}}
\caption{(Color Online) Electric field diagram of 4-electrode device simulated by COMSOL. The color shows the intensities of electric fields along the growth direction (a-d) or molecular axis (e, f). The electric field lines are shown in black while the arrows show the direction and magnitude of the electric field at the level of the LQDMs. The right column shows a zoom-in view inside the white rectangular zone on the left column. The pink profiles in the figures in the right column show the cross-sectional view of a single LQDM located inside the white rectangle in the left column. \label{model}}
\end{figure*}

To meet the design criteria for electric fields in both vertical and lateral directions, we turn to a four-electrode device. Conceptually, one pair of electrodes applies a lateral electric field while the other applies a vertical electric field. By combining the two electric fields, we are able to apply a two-dimensional vector electric field to a lateral or vertical QDM at an arbitrary location within the device. The device design also allows the isolation and controlled optical interaction with a single QD or QDM. Fig.~\ref{4ele} shows a schematic three-dimensional view of the device incorporating a LQDM sample. The distance between the QDM and the doped layer must be sufficiently thin that electrons can tunnel into the device to charge the QDM but sufficiently thick that sequential charging by one electron at a time can be controlled. The distance between the QDM and the sample surface impacts the change in field required to step through charging events; a larger distance makes discrete charging robust against small fluctuations in the electric field. Initial parameters for these thickness values were set based on our previous experience with QDM structures applying electric fields only in the growth direction. The device parameters we report below were obtained by iteratively optimizing to maintain control of charging while maximizing the electric fields applied at the QDM under fixed voltages.

Device fabrication starts with molecular beam epitaxy growth of a heterostructure incorporating LQDMs, following methods described elsewhere.\cite{Lee2010, Zhou2011} The molecular axis of the LQDMs naturally orients along the [0 1 -1] axis of the GaAs crystal due to anisotropic diffusion. The distance from the QDMs to the doped layer is 120 nm and the distance between the QDMs and the top surface of the sample is 330 nm. A mesa 1 $\mu$m wide and 370 nm deep is then etched so that the molecular axis of the LQDMs is perpendicular to the long axis of the mesa. The mesa is coated with a thin layer of $Al_{2}O_{3}$ (20 nm) that prevents charging of the LQDMs from the lateral electrodes and reduces the leakage current. We expect that atomic layer deposition (ALD) will be required for this step because ALD-grown $Al_{2}O_{3}$ is known to effectively passivate surface states and unpin the Fermi level in GaAs-oxide diodes\cite{Ye2003APL, Xuan2007}. A pair of lateral electrodes including 8 nm Ti and 100 nm Al are then deposited on the sides of the mesa. The thin Ti layer improves the adhesion of the metal electrodes to the GaAs. The electrode feature is defined by lithography followed by lift-off or dry etching to open up the top of the mesa and avoid shorts. Ideally the electrodes would terminate at the sides of the mesa to apply a purely lateral electric field, but this approach requires high precision in the layer-to-layer alignment. To relax the processing requirements and ensure a symmetrical structure, we choose a design that allows the electrodes to cover the side walls and top edges of the mesas. The gap between the two lateral electrodes is designed to be 800 nm, 200 nm smaller than the width of the mesa. Bond pads for connection to an external circuit are included in this layer.

Following the deposition of the lateral electrodes, a $SiO_{2}$ insulating layer with a thickness of 300 nm is deposited to cover the mesa and electrode fingers. The vertical electrode and its bonding pad are deposited on top of the $SiO_{2}$ insulating layer. The top electrode is composed of a semi-transparent Ti layer (8 nm) completely covering the mesa and a thicker Al layer ($\geq$ 100 nm) with 1 $\mu$m gaps oriented perpendicular to the mesa. This Al mask isolate sections of the mesa for optical measurements of individual LQDMs without needing to align the features during fabrication. The Ti layer guarantees the uniformity of electric fields along the growth direction. As shown on the top of Fig.~\ref{4ele}, the bonding pads for lateral electrodes are not covered by the insulator layer or subsequent features.

Fig.~\ref{model} presents device simulations that illustrate the capacity of this device to apply vector electric fields to a LQDM. In order to demonstrate the tolerance of this device to the arbitrary location of the QDMs, we randomly assume the location of the LQDM is 200 nm left of the center of the mesa. In Fig.~\ref{model} (a, b), a quasi-vertical electric field is applied to the region of the LQDM. In Fig.~\ref{model} (c, d), a vector electric field containing both lateral and vertical parts is applied. In Fig.~\ref{model} (e, f), a quasi-lateral electric field is applied. We note that this 4-electrode design is able to apply a maximum lateral electric field 1.2 to 2 times larger than the 3-electrode device. To clearly show the uniformity of the electric field magnitude and direction across the LQDM when a quasi-lateral field is applied, we choose a narrower range for the color scale in Fig.~\ref{model}(F). The fiigure shows that the lateral field varies from -1 to -1.5 MV/m from the left to the right of the LQDM. This is not a perfectly uniform electric field, but because the direction of the field is parallel to the molecular axis small variations are not likely to substantively alter the ability to tune the relative energy levels of the two QDs, which is the primary function of this lateral field.

In summary, we have presented a design for a four-electrode device that can apply vector electric fields to single QDs or QDMs located at arbitrary locations within a heterostructure mesa. We validate and illustrate the device performance with numerical simulations. We demonstrated the importance of the fourth electrode/ mask by comparing the simulation results to a simpler 3-terminal device design. The 4-electrode device presented here provides a feasible approach to achieving new functionality for optoelectronic devices based on QDs and will facilitate the emergence of quantum device technologies.

The authors acknowledge Juejun Hu and Hongtao Lin for providing simulation tools and instructions. We thank Xiaohong Yang and Antonio Badolato for helpful discussions. This research is supported by National Science Foundation grant No. DMR-0844747.

%\bibliography{QDLib_diode}

\end{document}